# Highly efficient light extraction and directional emission from diamond color centers using planar Yagi-Uda antennas

**Hossam Galal[1,*], Assegid M. Flatae[1], Stefano Lagomarsino[1,2], Gregor Schulte[3], Christoph Wild[4], Eckhard Wörner[4], Nicla Gelli[2], Silvio Sciortino[2,5,6], Holger Schönherr[3], Lorenzo Giuntini[2,5], Mario Agio[1,6]**

1. Laboratory of Nano-Optics and Cμ, University of Siegen, 57072 Siegen, Germany
2. Istituto Nazionale di Fisica Nucleare, Sezione di Firenze, 50019 Sesto Fiorentino, Italy
3. Laboratory of Physical Chemistry I and Cμ, University of Siegen, 57072 Siegen, Germany
4. Diamond Materials GmbH, 79108 Freiburg, Germany
5. Department of Physics and Astronomy, University of Florence, 50019 Sesto Fiorentino, Italy
6. National Institute of Optics (INO-CNR), 50125 Florence, Italy

*hossam.galal@uni-siegen.de

Color centers in diamond represent a promising platform for developing solid-state single-photon sources and spin-photon interfaces as building blocks for photonics-based quantum technologies. However, although they exhibit a combination of features that make them so attractive, they also suffer from limited control in their emission properties, such as brightness, directionality, and polarization to cite a few. In this paper we present implementations and the experimental investigation of planar Yagi-Uda antennas in diamond, demonstrating highly efficient light extraction and directional emission from silicon-vacancy color centers created in thin diamond membranes.

**Introduction**

Photonics-based quantum technologies require solid-state platforms that are capable of performing a number of operations on qubits [1,2,3]. Diamond has been recognized as a highly promising material system for this purpose, essentially because its color centers offer the possibility to implement such tasks at room temperature [4]. The fact that a high degree of control over the photon-matter interface is needed for achieving the desired performances has initiated a huge effort on the development of photonics strategies for improving light-matter interaction with diamond color centers [5,6]. Specific examples include photonic crystals structures [7], solid-immersion lenses [8], antennas [9,10] and scanning microcavities [11]. While these efforts are effective, they require sophisticated nanofabrication techniques and complex experimental approaches. Following the introduction of planar directional antennas, or planar Yagi-Uda antennas, which improve the directional emission of single molecules [12], we have recently reported on the potential of circumventing the degraded light extraction efficiency and the poor directional emission for emitters embedded in high-refractive index media; namely, diamond and semiconductors [13]. The scheme relies on the incorporation of low-refractive index layers in a planar form of a Yagi-Uda antenna [12]. Moreover, the layers raise the constraints on the positioning of the light emitters within the host medium, the layer hosting the emitters, making light extraction and emission directionality almost independent of the emitters' location. Thus, offering a larger tolerance margin to accommodate technological limitations.

Here we highlight how with the inclusion of tens to hundreds of nanometers of silica, a low-refractive index transparent material compatible with many industrial platforms, the emission profile of Silicon Vacancy (SiV) color centers in diamond [14] can be significantly enhanced. The uniqueness of the approach remains in its simplicity, with a couple of thin flat layers stacked; and its applicability to various materials and light emitters at different wavelengths.

**Theory**

The foreseen enhancement, basically, is an exploitation of the first order mode of the planar Yagi-Uda antenna, where the thickness of the host medium in correlation with the thicknesses of the low-refractive index layers have to be quite small. For some applications; for example: Light Emitting Diodes (LEDs) and lasers; more space can be required between the two antenna metal elements (reflector and director) for the incorporation of supporting layers, such as charge or transport layers, and in this case, one can take advantage of the antenna's higher order modes. These modes, in fact, are more promising if higher emission directionality is favorable. On the other hand, their light extraction efficiencies are inferior to their first order counterparts. The reason being is that; thicker active media, the overall layers bounded between the antenna's two metallic elements, beyond a certain thickness threshold support guided modes and trap a considerable amount of the emitted light [1]. However, with the knowledge of the structure's dispersion; one can make the coupling to these guided modes minimal by tuning the thickness of the individual active medium layers.

As a start, we briefly review the general antenna architecture shown schematically in Fig. 1a, which consists of metallic reflector and director layers and of low-refractive index layers to space the host medium from the antenna's metallic elements; with an additional low-refractive index layer directly on top of the director. The intermediate layers and the top layer are independent of each other, and can be used separately. However, together they can push the overall light emission efficiency to unprecedented values [13]. In this report, we demonstrate the antenna's potential in light emission enhancement from color centers, without emphasizing on driving its performance towards the peak limit.

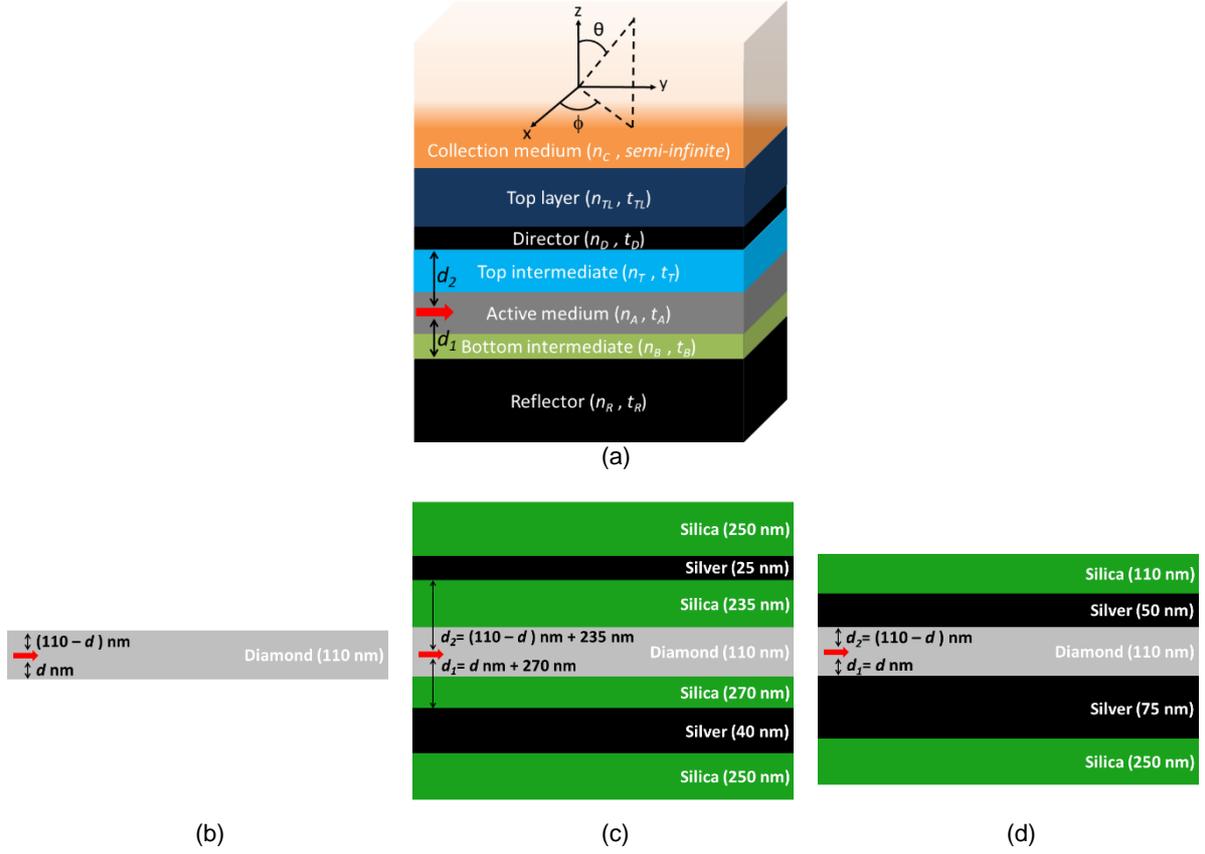

Fig. 1 (a) Advanced configuration of the planer Yagi-Uda antenna (reflector, bottom intermediate layer, top intermediate layer, director, top layer). Active medium hosting a horizontally oriented Hertzian dipole emitter (red arrow). The emitter is positioned a distance ($d_1$) from the reflector and a distance ($d_2$) from the director. (b) Reference design (Ref): a bare 110 nm thick diamond membrane, with a SiV emitter modeled as a horizontally oriented Hertzian dipole (red arrow), positioned a distance (*d*) from the bottom diamond interface. (c) Antenna design 1 (Ant1): silica top and bottom intermediate layers, and silver reflector and director; have been applied to the bare membrane Ref. The silica layers on top of the director and beneath the reflector are for the protection of the silver layers, and not part of the antenna's actual design. (d) Antenna design 2 (Ant2): silver reflector and director have been applied directly to the bare membrane Ref without intermediate layers. A top layer from silica has been used. This layer also serves for the protection of the silver layer, together with the silica layer beneath the reflector.

We verify the scheme on 110 nm thick diamond membranes, designated as Reference (Ref), see Fig. 1b; in two selective antenna configurations, Antenna design 1 (Ant1) and Antenna design 2 (Ant2), as depicted in Fig. 1c and 1d respectively. Throughout our calculations, we consider the SiVs as horizontally oriented Hertzian dipoles, positioned in the diamond membrane at *d*=30 nm within a ± 25 nm pitch. For the antenna configurations, the dipole's distance from the reflector ($d_1$) and the director ($d_2$) fulfill the relationship: $p\lambda/(6n) < d_1, d_2 < p\lambda/(4n)$; where λ is the emission wavelength, *n* is the refractive index of the active medium, and *p* is the mode order. The design of Ant1 is within the higher order range, with *p*=3; while design Ant2 exploits the first order mode, with *p*=1. The refractive indices of diamond, silica, and silver at 738 nm read: 2.4, 1.45, and 0.033+5.1i, respectively [15]. The collection and the background media are considered to be air, and the fluorescence is collected from the director side. Results are based on a plane-wave expansion of the dipole's emission in a multilayer structure [16,17].

Designs Ant1 and Ant2 are not the advanced (complete) version of the antenna scheme. In Ant1, Fig. 1c, only intermediate layers are incorporated; and the two 250 nm silica layers on top of the director and below the reflector are solely for technological reasons, to protect the silver layers from oxidation. While in Ant2, Fig. 1d, intermediate layers are not used; instead, the thickness of the top protective silica layer has been reduced to 110 nm. This layer not only protects the silver director but also serves as the top layer, and in conjunction with the collection medium tunes the dispersion of the director [13]. Ideally, the thickness of the reflector should be in bulk range, but in the two designs, it has been slightly reduced to allow for some transparency, in case of optically pumping the emitters from that side. These amendments, whether the protective layers or the reduction in the reflector's thickness, have no impact on the antenna's performance.

In assessment of the two different antenna designs, we recall two figures of merit: the outcoupling efficiency (β) and the beaming efficiency (η) [13]. β is given by $P_{rad}/P_{tot}$ : where ($P_{tot}$) is the total power emitted by the dipole in the layered structure, and ($P_{rad}$) is the power transmitted to the collection medium; both in terms of power densities as a function of the in-plane wavevector ($k_p$). By calculating $P_{rad}$ as a function of the azimuthal and polar angles, ϕ and θ respectively (see Fig. 1a), η is given by the ratio between the power collected up to a semi-angle θ of 20° and $P_{rad}$ ; corresponding to the power collected by a high-Numerical Aperture (NA) single-mode fiber (NA ~ 0.34). Ant1 features a very high directional emission profile, with η approaching 90% and the emission pattern funneled into a single-lobe having a radiation Half Angle (HA) at Full-Width Half Maximum (FWHM) as tight as 8°; in conjunction, β shows values around 10%. On the other hand, design Ant2 exhibits a higher extraction efficiency, with β in the range 40-60%; and η in the vicinity of 62% with a 15° HA single-lobe. Overall, and in comparison with Ref, both Ant1 and Ant2 show a significant enhancement in the measured fluorescence signal; whether translating from the strong light extraction or beaming effect, or the combination of both.

**Sample fabrication**

A cross-section schematic of the fabricated antenna samples, in particular, Ant1, is shown in Fig. 2a. The different color layers are in correspondence with that in Fig. 1c; however, the scaling of the antenna elements in both figures are not informative. The fabrication process starts with the growth of the polycrystalline diamond membrane (grey) on a silicon substrate (yellow); owing to the latter's flat surface. The membrane's bottom surface is then exposed by etching a hole thoroughly through the silicon substrate, via Reactive Ion Etching (RIE), leaving the diamond membrane suspended in air. SiV color centers are then created in a controlled manner through a $5 \times 10^7 cm^{-2}$ Si ion fluence implantation, followed by thermal annealing [18]. The silica (green) and silver (black) layers are then coated using an Edwards electron beam evaporator. One side of the antenna is coated at a time. The sample is left to cool down before it can be flipped to coat the other side of the antenna.

After applying the complete antenna structure to the suspended diamond membrane, the overall multilayer stack takes a curvature as shown in Fig. 2a. This bending applies to both Ant1 and Ant2, and presumably, is a resultant of the asymmetry in the deposited layers on both sides of the membrane, and due to the discrepancy in thermal constants of the individual materials; especially that a cool down period interposed the coating of the two antenna sides. One has to carefully control temperature gradients to avoid snapping the membrane.

## Results and discussions

The three samples Ref, Ant1, and Ant2 were optically characterized using a 656 nm laser, in a home-built confocal microscopy setup as depicted schematically in Fig. 2b. The fluorescence signal is collected by means of an air objective lens, and undergoes necessary filtering before detection on the instruments.

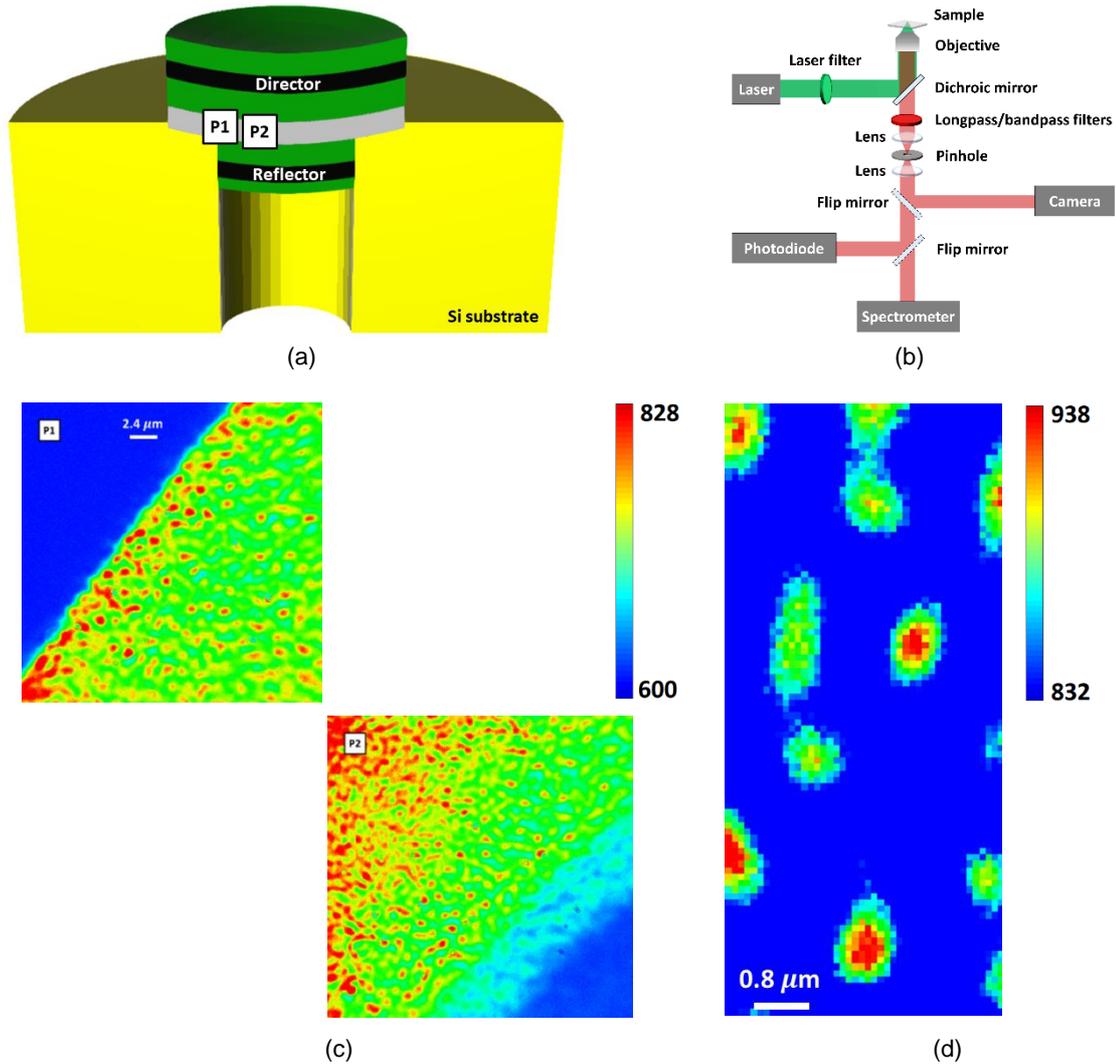

Fig. 2 (a) Sample Ant1 cross-section schematic. The diamond membrane (grey) is suspended on an annular silicon substrate (yellow). Silica layers (green) and silver layers (black) are deposited on the suspended membrane to construct the design shown in Fig. 1c. Points (P1 and P2) mark roughly the locations on the sample where the images in (c) were taken. (b) Schematic of the home-built confocal microscope used in the optical characterization of the samples. Instrument specifications: [laser (PicoQuant, PDL 800-D, LDH-D-C-660); camera (Princeton Instruments, ProEM-HS: 512 BX3, back-illuminated electron multiplying CCD (EMCCD) camera); spectrometer (Andor, Shamrock 500i) equipped with an EMCCD camera (Andor, Newton 970, A-DU970P-BVF); photodiode (APD) (Micro Photon Devices, 50 cps dark count, < 50 ps jitter); and a Time-Correlated Single Photon Counter (TCSPC) (PicoQuant, PicoHarp 300). (c) Wide-field fluorescence images for sample Ant1, taken at the spots (P1 and P2) marked in (a). Fluorescence from the SiVs appears in red-yellow mapping, in both images. Blue color map in (P1) represents the edge of the silicon substrate, while in (P2) represents defocused planes; with the SiVs' fluorescence also appearing in green gradient mapping near the defocused planes, in (P2). (d) A more focused fluorescence image for sample Ant1, showing an ensemble of eight emitters with varying brightness (red: in focus; green out of focus).

**Fluorescence imaging**

The bending downwards, reported earlier in Fig. 2a, is also evident in the fluorescence images of Fig. 2c; measured for design Ant1 with a 0.75NA-100x ZEISS objective. The two wide-field images in Fig. 2c were taken on the two spots (P1 and P2) marked in Fig. 2a. The Zero Phonon Line (ZPL) emission from the SiVs, appears in red color map, in both images. At (P1), the edge of the silicon substrate blocks the light, and therefore appears in blue; while by moving towards the center of the membrane, at (P2), the region close to the center of the membrane appears in blue due to the defocusing of its plane. One has to readjust the depth position of the objective when moving towards the center of the membrane, to observe the emitters. A more focused image view of an emitter ensemble, also from design Ant1, is shown in Fig. 2d.

**Back focal plane imaging**

Angular information on the emitters' radiation profile has been extracted from the Fourier plane, using Back Focal Plane (BFP) imaging. Theoretical calculations expect a single-lobe radiation pattern for Ref, Ant1, and Ant2; however, with significant distinctions in directionality: HA≈40º and $\eta$≈12%; HA≈8º and $\eta$≈86%; HA≈15º and $\eta$≈62%; respectively. In case of Ref, the radiation HA shows strong dipole-position dependency; while this is not the case for both antenna designs. The three sub-figure columns in Fig. 3 are in correspondence with Fig. 1b,c,d; Ref, Ant1, and Ant2 respectively. The first sub-figure row in Fig. 3 is the theoretical far-field power radiation pattern; followed by the BFP images of selected emitters, in the second row; and comparisons between theory and experiment in terms of the radiated power and the radiation angles are drawn, along the last row. Color map scaling in Fig. 3d,e,f are only for clarification, and should not be confused and used to compare the brightness of the emitters. The pixel scaling in the BFP images for the three samples is not the same. Neighboring emitters in Fig 3e,f have been masked out, for the demystification of the results.

In spite of the distribution in the emitters' position, due to the statistical nature of the implantation process, good agreement between theory and experiment can be overall remarked in Fig 3; which emphasizes on the antenna's robustness in accommodating technological imperfections.

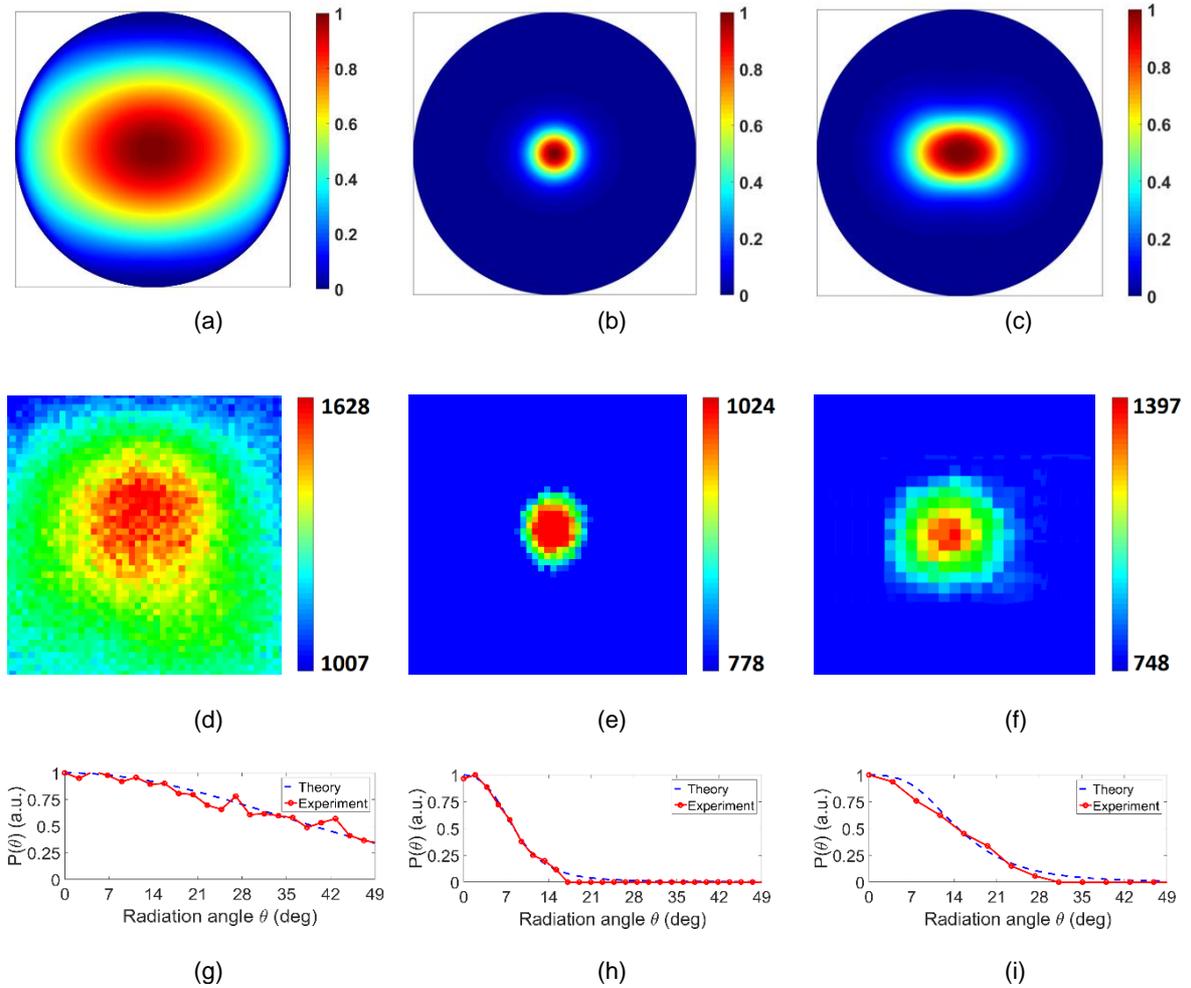

Fig. 3 (a), (b), and (c) Theoretical far-field power radiation pattern, on a normalized scale; for Ref, Ant1, and Ant2 respectively. (d), (e), and (f) BFP images for the radiation pattern of Ref (scale: 46x46 px, 2.25°/px), Ant1 (scale: 54x54 px,1.9°/px), and Ant2 (scale: 26x26 px, 3.9°/px) respectively. Color mapping scales in the three BFP images are not informative on the emitters' brightness. Neighboring emitters in (e) and (f) masked out. (g), (h), and (i) The radiation half angle θ versus the normalized radiated power, theory vs experiment, for Ref, Ant1, and Ant2 respectively.

### Fluorescence count rate

Here we investigate the overall enhancement in emission count rate, namely β×η, for Ant1 and Ant2 in two different types of measurements; shown in Fig. 4a,b respectively. The samples are excited from the director side, and the fluorescence signal is collected with the same objective. In the first configuration, the laser power ($P_{Laser}$) used to excite Ref and Ant1 is the same; however, the pump power ($P_{Pump}$) for Ant1 is much weaker than that for Ref, owing to the director's high reflectivity (R~0.89). While in the second configuration, $P_{Laser}$ is adjusted to maintain the same $P_{Pump}$ for both Ref and Ant2. The emission count rate from Ref is already very weak, especially with a low $P_{Pump}$, and therefore a higher numerical aperture objective has been utilized in the latter configuration. Reminding that, the antenna works only for emitters with Horizontal (H) dipole orientations. Vertical (V) dipoles will be completely quenched, and only the in-plane component of Diagonal (D) dipoles will survive, as illustrated in Fig. 4a,b [12,13]. The polycrystalline nature of the diamond membrane does not give preference to a certain dipole orientation. Statistically, we can assume a one-third probability of occurrence for each dipole orientation, H, V, and D. More emitters will be effectively active in case of Ref, and accordingly, one would expect a higher count rate than for Ant1 and Ant2.

The spectral emission count rate of Ant1 versus Ref are plotted in Fig. 4c, on normalized scale; in observation of the antenna's beaming effect dispersion. In Fig. 4d, Ant2 and Ref are plotted with their actual count scales, for emphasis on the discrepancies in count rates. With Ref in red and Ant1 in blue, Fig. 4c, one can already remark the submersion of Ref''s 738 nm ZPL in a broad background emission. This background originates from impurities incorporated during the growth process of the membranes, and due to the polycrystalline nature of the membranes. Along the full spectral range, three different behaviors stand out with respect to Ant1's response, indicated with the black arrows: the short-wavelength range background emission is enhanced; the long-wavelength range background is quenched; and the ZPL is ~5 nm blue shifted. The background modulation has been studied in [13]. This collective behavior translates from the broadband operation of the antenna; the beaming effect already starts earlier before the ZPL, peaks at the ZPL, and then ramps down onwards. As for the ZPL spectral shift, the membrane bending reported earlier is a result of the strain build up within the membrane. The different materials deposited on the membrane exert a tensile/compressive force on the membrane. With the membrane suspended in air, it relaxes to the strain with a slight curvature, and takes the shape of a meniscus structure, similar to that in Fig. 2b. Near the edges of this structure, the curvature (strain) is maximum; while the curvature (strain) is minimal towards the center. By sampling measurements along the in-plane of the membrane, and in moving from the edge to the center of the membrane; Ant1's ZPL experiences a maximum blue shift near the edge, cyan plot Fig. 4e, and converges (red shifts) to Ref's ZPL (red dashed plot) in proximity to the center (along the black arrow). Similar spectral shifts have been reported in SiVs, upon the application of an external strain in a controlled manner [19,20].

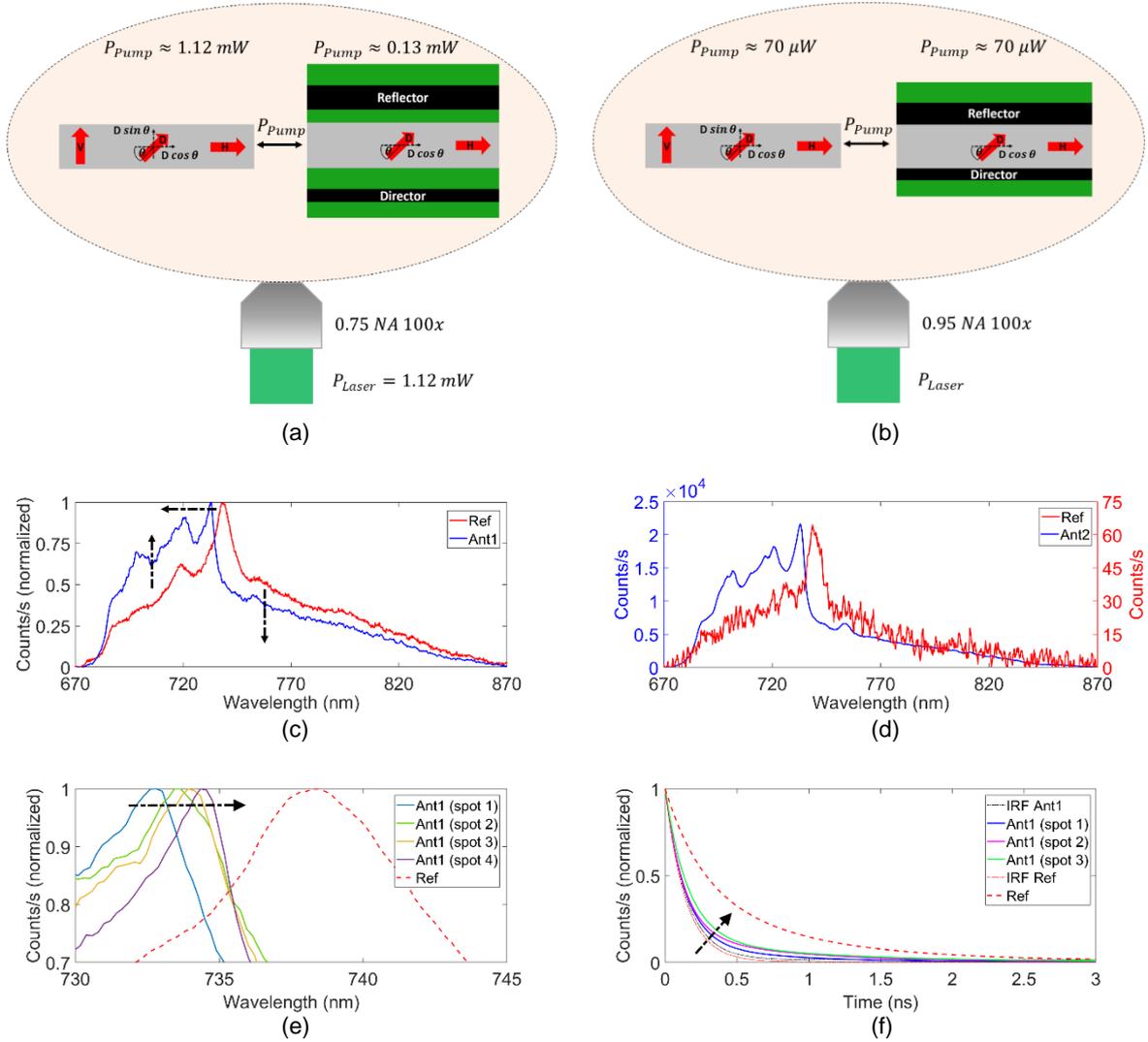

Fig. 4 (a) Experimental aspects of the measurements performed individually on Ref and Ant1 (same $P_{Laser}$), 0.75NA-100x ZEISS objective used. The corresponding results are shown in (c). (b) Experimental aspects of the measurements performed individually on Ref and Ant2 (same $P_{Pump}$), 0.95NA-100x Olympus objective used. The corresponding results are shown in (d). (c) Count rate spectral response for Ref vs Ant1, on normalized scale. Vertical black arrows indicate the background modulation for Ant1 with respect to Ref; horizontal black arrow indicates the blue shifting of the ZPL of Ant1. (d) Count rate spectral response for Ref vs Ant2, on actual scale. The response of Ref is projected on the right y-axis, while the response of Ant2 is projected on the left y-axis. (e) ZPL spectral shift measurements for Ant1; performed by measuring four spots, starting from the edge of the membrane (spot 1) and by moving towards the center of the membrane (spot 4). The ZPL for Ref is in red dashed plot. Black arrow indicates the direction towards the center of the membrane. (f) Lifetime measurements for Ant1; performed by measuring three spots, starting from the edge of the membrane (spot 1) and by moving towards the center of the membrane (spot 3). The Ref lifetime is in red dashed plot, corresponding instrument response functions are in dotted plots. Black arrow indicates the direction towards the center of the membrane.

The aforementioned inhomogeneous strain argument has also been registered in lifetime measurements, for Ant1. However, the intention here is not to study the lifetime dynamics, but to visualize the spectral shift from different perspectives. The lifetime of Ref and Ant1 were measured at different times, with the corresponding Instrument

Response Function (IRF) of each marked in dotted plots, shown in Fig. 4f. For Ant1, the lifetime is shortest at the membrane's far edge, blue plot, and by moving towards the center of the membrane (following the arrow) it converges towards that for Ref, red dashed plot. Additional band-pass filters have been utilized to filter out the background emission.

Coming back to the count rate measurements for Ant2, Fig. 4b; the same $P_{Pump}$ allows us to assess the enhancement in count rate on a one-to-one basis. Again, we stress on the point that, more emitters are contributing to the overall count rate in case of Ref. The high numerical aperture objective used specifically in this measurement does not favor the photon collection for Ant2; because, already its emission profile is highly directional. In fact, the radiation angles in case of Ref are very broad, and a high NA objective will certainly account for this divergence and deliver a higher count rate. In practice, we can extend the 0.34 NA collection efficiency $\eta$ to a 0.75 NA and a 0.95 NA objective lens, $\eta_{0.75}$ and $\eta_{0.95}$ respectively. In terms of numbers: $\eta_{0.75} \approx 54\%$ and $\eta_{0.95} \approx 90\%$, for Ref; $\eta_{0.75} = 96\%$ and $\eta_{0.95} = 100\%$, for Ant2. The results are shown in Fig. 4d, with the response of Ref in red projected on the right y-axis; and Ant2 in blue projected on the left y-axis. The Ref ZPL registers a count rate of ~60 Cps, at the best. On the other side, the blue shifted Ant2 ZPL registers a count rate of more than 21000 Cps. This corresponds to an enhancement by more than 333 times, with the same $P_{Pump}$ and with a fewer number of emitters. It is also worth mentioning; in spite that Ref was pumped ~9 times harder than Ant1 in the experimental configuration of Fig. 4a, and with a fewer number of emitters, Ant1 still shows ~3 times higher count rate than Ref.

There have been efforts to characterize a single emitter for the three samples. However; unlike single crystalline diamond, the bare polycrystalline diamond membranes inherently exhibit a strong background emission superimposed on the ZPL. With the application of the antenna, the background is strongly amplified and diminishes the observation of a dip in the second order correlation measurements.

**Conclusion**

We have demonstrated a simple strategy to overcome the poor light emission brightness in diamond color centers, in a completely flat configuration; and just with a couple of ultrathin layers, five at maximum, the light extraction and the directionality efficiencies can simultaneously be brought to unprecedented values. The flexibility of the scheme, offering different operational mode regimes; the applicability to different materials at different wavelengths, e.g. to the nitrogen vacancy center; and the robustness against technological imperfections, hold promises for quantum technologies and also for many light emission and detection applications, such as LEDs, biosensors and IR detectors.